\documentclass[printer]{aa}

\usepackage{graphicx} 
\usepackage{txfonts} 
\begin{document} 
\title{An abundance analysis of the symbiotic star CH~Cyg} 
 
\author{M.R. Schmidt \inst{1} 
       \and 
        L. Za\v{c}s \inst{2} 
       \and 
        J. Miko\l ajewska \inst{3} 
       \and 
        K.H. Hinkle \inst{4} 
       } 
 
\offprints {M.R.Schmidt} 
 
\institute{N.Copernicus Astronomical Center, ul. Rabia\'nska 8, 
           PL-87-100 Toru\'n, Poland \\ 
           \email{schmidt@ncac.torun.pl} 
     \and 
           Institute of Atomic Physics and Spectroscopy, 
           University of Latvia, Rai\c{n}a bulv\= aris 19, 
           LV-1586 R\=\i ga, Latvia \\ 
           \email{zacs@latnet.lv} 
     \and 
           N.Copernicus Astronomical Center, ul. Bartycka 18, 
           PL-00716 Warsaw, Poland  \\ 
           \email{mikolaj@camk.edu.pl} 
     \and 
           National Optical Astronomy Observatory, 
           P.O. Box 26732, Tucson, AZ 85726, USA \\ 
           \email{hinkle@noao.edu} 
          } 
 
\date{Received; accepted } 
 
\titlerunning{Abundance analysis of the symbiotic star CH~Cyg}

\authorrunning{M.R. Schmidt et al.}

\abstract{The photospheric abundances for the cool component of the
symbiotic star CH~Cyg were calculated for the first time using
high-resolution near-infrared spectra and the method of 
of standard LTE analysis and atmospheric models.  The
iron abundance for CH~Cyg was found to be solar, [Fe/H] =
0.0$\pm$0.19.  The atmospheric parameters ($T_{\rm eff}$ = 3100\,K,
$\log g = 0.0$ (cgs), $\xi_{\rm t} = 2.2$ km\,s$^{-1}$) and metallicity
for CH~Cyg are found to be approximately equal to those for nearby
field M7 giants. The calculated [C/H] = -0.15, [N/H] = +0.16, [O/H] =
-0.07, and the isotopic ratios of $^{12}$C/$^{13}$C and
$^{16}$O/$^{17}$O are close to the mean values for single M giants that
have experienced the first dredge-up. A reasonable explanation for the
absence of barium star-like chemical peculiarities seems to be the high
metallicity of CH~Cyg. The emission line technique was explored for
estimating CNO ratios in the wind of the giant. 
%
%
 
} 
 
\maketitle

\section{Introduction} 
 
Both symbiotic and peculiar red giants (barium, CH-stars, extrinsic S
stars, etc.) are binaries with a cool giant primary and a hot (WD)
secondary.  Interaction between the binary components results in
peculiar chemical composition in the atmospheres of barium and related
giants.  This is believed to result from the previous transfer of
$s$-process rich mass from the AGB star, which is now a WD, to the star
that now appears as the red giant (Han et al., \cite{han95}; Za\v{c}s
\cite{zacs}).  Symbiotic activity indicates that these binaries are
currently interacting.  Symbiotic systems also have a white dwarf that
was previously an AGB star and a red giant that was a main sequence
star during the AGB phase of the now white dwarf.  An obvious
expectation is that there are red giants with WD companions that are
both chemically peculiar and symbiotic (see, for review, Jorissen
\cite{jorissen03}).  Although a number of symbiotic stars are well
studied there is a lack of detailed abundance analysis based on
high-resolution spectra.

The link between the $yellow$ symbiotics and peculiar giants has been
established using the abundance patterns (Smith et al. \cite{smith96},
\cite{smith97}; Pereira, Smith, \& Cunha \cite{pereira98}; Smith,
Pereira, \& Cunha \cite{smith01}). All known yellow symbiotic stars
studied so far using atmospheric models display significant enhancement
of s-process elements, similar to peculiar giants (see, Jorissen
\cite{jorissen03}). They are clearly halo objects, as revealed by their
low metallicities and high space velocities.
The s-process Ba and Sr enhancements have been also found  in the two symbiotic stars containing a CH-type giant with a very high C overabundance (Schmid \cite{schmid94}). CH stars are also metal poor
halo objects.

For the $red$ symbiotics a link to the chemically peculiar giants is
much more tenuous. Only a small number of galactic red symbiotic stars
have a cool carbon-rich star as a cool component. Recently, peculiar
chemical composition was confirmed for the red symbiotic HD\,35155; the
star was found to be metal deficient, [Fe/H] = $-$0.8, and rich in
neutron-capture elements, [n/Fe] = + 2 dex on average (Vanture et al.
\cite{vanture03}). It is not clear why among red symbiotic stars the
percent of carbon and $s$-process rich stars is so low.
 
CH~Cyg (=HD\,182917 = HIP\,95413) is the brightest symbiotic binary at
visual wavelengths and the second brightest symbiotic in the 2 $\muup$m
infrared. The CH~Cyg system has been observed both photometrically and
spectroscopically from radio through $X$-ray wavelengths.  Both the
light curves and the radial velocity curves show multiple periodicites
(e.g. a $\sim 100^\mathrm{d}$ photometric period best visible in the
$VRI$ light curves, attributed to radial pulsations of the giant
(Miko{\l}ajewski, Miko{\l}ajewska \& Khudyakova \cite{mmik92}), and a
secondary period of $\sim 756^\mathrm{d}$ also present in the radial
velocity curve (Hinkle et al. \cite{hinkle93}). 

There is a controversy about whether the CH~Cyg system is triple or
binary system and if triple whether the symbiotic pair is the inner
binary or the white dwarf is on the longer orbit ((Hinkle et al.
\cite{hinkle93}; Munari et al. \cite{munari96}).  Although the
triple-star model is very appealing, this model does not easily explain
the observed photometric behaviour (Munari et al.  \cite{munari96}, and
references therein). Recent study of radial velocity variations of nine
multiple-mode semiregular M-giants (SRVs) revealed radial-velocity
periods longer than the fundamental radial mode in six of the nine
giants (Hinkle et al. \cite{hinkle02}). Although the authors considered
a possibility that the velocity variations are due to orbital motion,
the nearly identical orbital parameters ($K$, $e$, and to a lesser
extent also $\omega$) for five of the six SRVs make a binary model very
unlikely.  As discussed by Wood et al. \cite{wood04} and Hinkle et al.
\cite{hinkle02} evidence points to a stable non-radial pulsation of very
long period rather than duplicity.  The orbital parameters, $K$, $e$,
$\omega$, for the $\sim 756^\mathrm{d}$ pair of CH~Cyg (Hinkle et al.
\cite{hinkle93}) are practically the same as those for these five
multiple-period SRVs studied by Hinkle et al.  (\cite{hinkle02}).  This
strongly suggests that the $\sim 756^\mathrm{d}$ period in CH~Cyg is a
non-radial pulsation mode.  If so, the  $\sim 5300^\mathrm{d}$ velocity
variation discussed by Hinkle et al. \cite{hinkle93} must then be the
symbiotic orbit.

The cool giant has been classified as M7 (M6.5 - M7.5) by M{\"u}rset
\& Schmid (\cite{murset99}). In a quiet phase the cool component is a
typical oxygen-rich star with TiO bands dominating the red region of
the optical spectrum.  In the past CH~Cyg was used as a M6 spectral
standard (Keenan \cite{keenan63}). Chemical peculiarities are not
obvious in the optical spectrum of CH~Cyg.  However, due to the crowded
spectrum in the optical spectral region an analysis of chemical
composition is very difficult. Here for the first time we present a
detailed abundance analysis of CH~Cyg based on high-resolution,
near-infrared spectra, the method of standard LTE analysis and
atmospheric models, and spectrum synthesis. We have also looked for
variations of selected absorption lines.

\section{Observations and data reduction}

The high-resolution 2 $\muup$m region spectra of CH~Cyg were obtained
with the Fourier transform spectrometer (FTS) at the coud\'e focus of
the KPNO Mayall 4 m telescope (Hall et al. \cite{hall78}) between 1979
and 1992. The majority of the observations were taken with either a K
filter (2.0 - 2.5 $\muup$m) or a broadband 1.5 - 2.5 $\muup$m filter
covering both the K and H bands. The resolution of the spectra is 0.07
cm$^{-1}$ ($\sim$4~km\,s$^{-1}$). The typical signal-to-noise (S/N)
ratio of the spectra selected for abundance analysis range from 80 to
180 around 4600 cm$^{-1}$ and twice less in the H-band. For more
details about observations and reduction of the data see Hinkle et al.
(\cite{hinkle93}).

The six spectra of CH~Cyg were selected for further analysis in this
paper.  Five of them are in the K band where the spectrum is less
crowded and the continuum seems to be clearly defined. To cover
different phases of symbiotic activity the spectra observed on 11 June
1982, 30 December 1984, 14 March 1987, 15 May 1987, and 6 April 1991
were chosen.  In addition one spectrum observed on 13 January 1989 in H
band was selected for abundance analysis. All the selected spectra
have high S/N ratio and low level of fringing visible in some of the 46
spectra available.

\subsection{Identification and measurements of absorption lines} 

An inspection of the selected spectra of CH~Cyg shows that a large
number of lines for neutral atoms (Na, Mg, Al, Si, S, Ca, Sc, Ti, V,
Cr, Fe, Ni) and molecules (CO, CN, OH, HF) are evident. Also some lines
of H$_2$O were found near 4700 $cm^{-1}$ using the list of sunspot
water lines identified by Wallace et al. (\cite{WLH96}). 
To our knowledge no s-process element abundances have been studied in 
H and K bands.  The line databases used in the paper contain few
lines of Ba I and Y I in the spectral regions of interest with these lines
too weak to be identified.  Strong enhancements, $>$1 dex, would be
necessary to enable clear identification of these lines in the
spectrum.

Absorption line
identifications were made using  the list of atomic lines given by
M\'elendez \& Barbuy (\cite{melendez99}) in the H band, the VALD
database in the K band (Kupka et al. \cite{kupka99}) and the atlas of
cool stars spectra prepared by Wallace \& Hinkle (\cite{wallace96}).
The molecular lines were identified using the line list of CO and
isotopic vibration-rotation transitions given by Goorvitch \&
Chackerian (\cite{goorvitch1}), HF transitions prepared by Jorissen,
Smith, \& Lambert (\cite{jorissen92}), molecular data from Kurucz
(1995) and the  list of red system CN lines published by Davis \&
Phillips (\cite{davis63}).

\begin{figure} 
  \resizebox{\hsize}{!}{\includegraphics{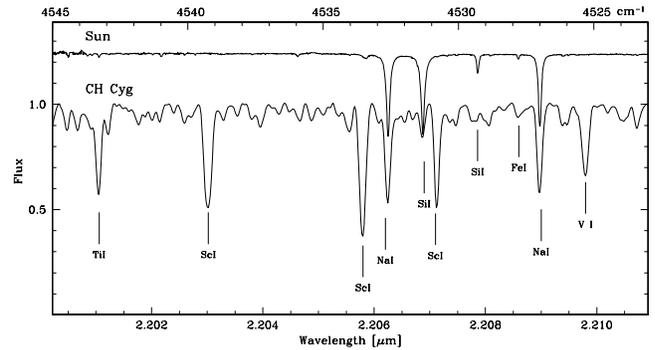}} 
  \caption{The observed spectrum of CH~Cyg in the region of  
well-known Ca index. Also shown is the spectrum of the  
Sun (Wallace et al. \cite{WLH96}).} 
  \label{fig1} 
\end{figure} 

\begin{figure} 
  \resizebox{\hsize}{!}{\includegraphics{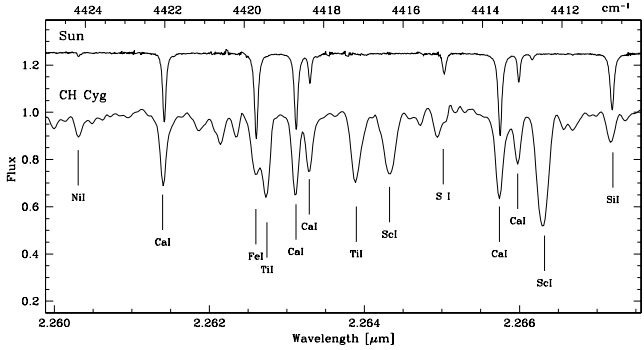}} 
    \caption{Same as Fig.1, but around the Na index.} 
     \label{fig2} 
\end{figure} 

No telluric reference spectra were taken for these CH~Cyg observations
and hence it was not possible to remove telluric lines from the
spectra. However, this prove not to be a serious limitation because
many stellar lines are uncontaminated by telluric features. Therefore
the first task was to identified stellar and telluric lines in the
observed spectra of CH~Cyg. An atmospheric transmission function
adopted from the Arcturus Atlas (Hinkle, Wallace, \& Livingston
\cite{hinkle95}) was used to select stellar lines whose positions
differ significantly from telluric features. In total more than 300
clean atomic and molecular lines for abundance analysis were selected,
however, only a part of them have acceptable spectral line data.
Equivalent widths (EW) of selected lines were measured by fitting with
the Gaussian using the standard DECH20T routine (Galazutdinov
\cite{galazutdinov92}) both in the spectra of CH~Cyg and the Sun.
Equivalent widths measured at different phases of stellar activity were
compared.  Only a small fraction of the equivalent widths displayed
significant changes, i.e. more than 10\%.  We are not sure that these
changes are due to physical effects. The variations in EW could be
because of uncertainties in the continuum definition and contamination
by neglected telluric components. Notice that the locations of telluric
lines change in the spectra with both heliocentric velocity and orbital
phase. Additional analysis is needed to understand nature of changes in
some EW. In this paper only constant stellar lines are exploited. In
the final list of lines for abundance analysis (see Tables 1 and 2)
clean absorption lines with acceptable atomic (molecular) data were
included and averaged equivalent widths are given.

Although normalization of spectra sometimes is difficult the abundance
calculations for iron over the observed spectral region gives evidence
that in general the continuum is clearly defined. Notice that the
spectrum of CH~Cyg in the H band is much more crowded than in the K
band. Therefore for abundance calculations we prefer the K band.  In
the K band it is also easier to detect fringing as a modulation of the
continuum.  The resulting rectified spectra for two typical wavelengths
regions are presented in Figs.~1 and 2.  Also shown is the spectrum of
the Sun (Wallace et al. \cite{WLH96}) which we used to check the
methodology of abundance calculations.

\begin{table*} 
\caption[]{$gf$-values, excitation potentials and averaged 
equivalent widths of atomic lines in the spectra of CH~Cyg and the 
Sun.} 
\label{TabElemCont} 
\begin{tabular}{cccccccccc} 
\hline 
Wavelength & EP &$\log gf$& Sun &CH~Cyg &Wavelength& EP&$\log gf$& Sun&CH~Cyg \\ 
(\AA)& (eV) &   & (m\AA)& (m\AA) & (\AA)& (eV)&    & (m\AA)&(m\AA) \\ 
\hline 
\ion{Mg}{i}&    &       &     &     & \ion{Fe}{i}&    &    &   &    \\ 
15952.74 & 6.59 & -1.98 &  8  & 228 & 16233.10 & 6.38 & -1.18  & 7  &  38  \\ 
16629.27 & 6.73 & -1.77 &     & 266 & 16247.52 & 6.28 & -1.09  & 14  &  54  \\ 
16636.57 & 6.73 & -1.51 &     & 224 & 16263.37 & 6.24 & -0.90  & 20  & 102  \\ 
17090.30 & 6.73 & -1.92 & 11  & 258 & 16281.95 & 6.32 & -0.51  & 38  & 113  \\ 
17754.47 & 6.73 & -1.75 &  8  & 210 & 16323.17 & 5.92 & -0.60  & 74  & 173  \\ 
17758.55 & 6.73 & -1.78 &     & 242 & 16441.12 & 5.92 & -0.56  & 82  & 156  \\ 
17758.60 & 6.73 & -1.47 &     & 242 & 16444.91 & 6.29 & -0.36  & 67  & 148  \\ 
17766.91 & 6.73 & -1.11 &     & 300 & 16520.18 & 5.56 & -2.41  & 3  &  48  \\ 
21065.48 & 6.78 & -1.55 &     & 213 & 16526.05 & 6.29 & -0.67  & 35  & 103  \\ 
21231.38 & 6.73 & -1.32 &  33 & 278 & 16656.78 & 6.34 & -0.65  & 34  & 125  \\ 
\ion{Si}{i}&    &       &     &     & 16683.73 & 5.92 & -1.05  & 34  & 164  \\ 
15642.74 & 6.73 & -1.76 &  10 &  35 & 16697.67 & 6.42 & -0.41  & 40  & 140  \\ 
15805.27 & 6.80 & -1.60 &  06 &  21 & 16726.03 & 6.38 & -0.52  & 36  & 166  \\ 
16064.41 & 5.95 & -0.66 & 279 & 239 & 16815.97 & 6.30 & -1.02  & 14  &  57  \\ 
16099.20 & 5.96 & -0.25 &     & 301 & 16904.85 & 6.30 & -1.27  & 9  &  45  \\ 
16384.61 & 5.86 & -1.00 & 194 & 246 & 16915.31 & 5.87 & -1.94  & 4  &  42  \\ 
17332.10 & 6.62 &  0.30 &     & 276 & 17025.41 & 5.07 & -2.48  & 6  & 110  \\ 
20922.86 & 6.73 &  0.69 & 418 & 280 & 17126.29 & 5.94 & -1.95  & 3  &  48  \\ 
20931.86 & 6.73 & -0.86 &  41 &  66 & 17493.34 & 6.41 & -0.78  & 28  &  96  \\ 
21360.03 & 6.22 &  0.08 & 408 & 302 & 17575.26 & 6.73 & -0.17  & 28  &  97  \\ 
21785.60 & 6.72 &  0.55 & 304 & 213 & 17586.72 & 6.38 & -0.55  & 42  &  90  \\ 
21825.63 & 6.72 &  0.26 & 235 & 218 & 17705.69 & 6.34 & -0.98  & 15  &  96  \\ 
21880.11 & 6.72 & -0.52 &  85 & 100 & 17719.21 & 6.58 & -0.64  & 14  &  96  \\ 
21885.30 & 6.72 &  0.52 & 311 & 220 & 20953.87 & 6.12 & -0.87  & 20  & 136  \\ 
\ion{Ti}{i}&    &       &     &     & 21183.92 & 3.02 & -4.17  & 9  & 237  \\ 
17022.68 & 4.51 &  0.37 &   4 & 168 & 21244.23 & 4.96 & -1.38  & 77  & 258  \\ 
17381.32 & 4.49 &  0.33 &   4 & 147 & 21290.17 & 3.07 & -4.37  & 4  & 244  \\ 
17387.85 & 4.47 &  0.23 &   4 & 153 & 21741.42 & 6.18 & -0.65  & 27  & 111  \\ 
17393.26 & 4.51 &  0.48 &   7 & 216 & 21819.72 & 5.85 & -1.58  & 7  &  63  \\ 
21158.08 & 4.86 &  0.52 &   2 & 124 & 21857.36 & 3.64 & -3.91  & 9  & 196  \\ 
21314.43 & 3.89 & -0.84 &   3 & 157 & 22391.13 & 5.32 & -1.53  & 23  & 146  \\ 
21872.05 & 3.92 & -0.62 &     & 163 & 22399.16 & 5.10 & -1.26  & 67  & 191  \\ 
\ion{Fe}{i}&    &       &     &     & 22426.14 & 6.22 & -0.17  & 62  & 139  \\ 
15643.22 & 5.81 &-1.81  &  7  & 33  & 22818.80 & 5.79 & -1.26  & 21  & 105  \\ 
15674.41 & 6.20 &-1.04  & 16  & 86  & 22852.17 & 5.83 & -0.67  & 69  & 224  \\ 
15675.28 & 6.33 &-0.57  & 32  & 69  &          &      &        & &      \\ 
15704.38 & 6.33 &-1.08  &  9  & 46  & \ion{Ni}{i}&    &        & &      \\ 
15756.01 & 6.36 &-0.85  & 11  & 63  & 16037.88 & 6.20 & -0.70  & 3  &  30  \\ 
15820.95 & 5.96 &-0.73  & 55  &120  & 16140.51 & 5.49 & -0.24  & 53  & 145  \\ 
15826.03 & 5.64 &-0.96  & 58  &161  & 16367.58 & 5.28 &  0.28    &165  & 310  \\ 
15844.52 & 6.36 &-0.40  & 46  & 83  & 16593.97 & 5.47 & -0.59  & 27  & 184  \\ 
15862.99 & 5.59 &-1.34  & 29  &129  & 16710.63 & 6.03 & -0.97  & 2  &  41  \\ 
15900.89 & 6.34 &-0.89  & 12  & 95  & 16820.06 & 5.30 & -0.70  & 29  & 205  \\ 
15933.83 & 6.31 &-0.59  & 33  &116  & 17005.67 & 5.49 &  0.15    &107  & 226  \\ 
15956.99 & 6.34 &-0.81  & 26  & 99  & 17311.29 & 5.49 & -0.63  & 22  & 169  \\ 
16017.24 & 5.59 &-1.97  &  8  & 56  & 17365.45 & 6.07 & -0.29  & 13  & 109  \\ 
16033.80 & 6.35 &-0.63  & 31  & 75  & 17898.76 & 6.10 & -0.27  & &  86  \\ 
16176.34 & 6.38 &-0.52  & 33  &113  & 17993.39 & 6.10 &  0.01  & & 161  \\ 
16182.44 & 6.38 &-0.52  & 34  & 85  & 21575.89 & 5.30 & -1.32  & 10  & 113  \\ 
\hline 
\end{tabular} 
\end{table*} 

\begin{table*} 
\caption[]{$gf$-values, excitation potentials and equivalent 
widths of molecular lines in the spectrum of CH~Cyg.} 
\label{TabMol} 
\begin{tabular}{cccclccccc} 
\hline 
Wavelength &$\chi$& $\log gf$  & EW &Line & Wavelength &$\chi$& $\log gf$&  EW &Line \\ 
(\AA) & (eV) &   & (m\AA) &    & (\AA) & (eV) &   & (m\AA) &    \\ 
\hline 
  OH  &      &       &       &       &   CO  &     &    &     &  \\ 
17054.7187 & 1.10  &   -4.65  &     394   & (4-2)\,P$_{2}$-9.5  & 
16285.845  & 1.5986 &   -5.7660 &   183    & (6-3)\,R59 \\ 
17056.8417 & 1.10  &   -4.65  &     409   & (4-2)\,P$_{2}$+9.5  & 
16300.055  & 0.5707 &   -7.1392 &   270    & (5-2)\,P13 \\ 
17074.1425 & 1.10  &   -4.65  &     346   & (4-2)\,P$_{1}$-10.5 & 
16301.731  & 1.6533 &   -5.7402 &   261    & (6-3)\,R61 \\ 
17099.2187 & 0.84  &   -5.01  &     373   & (2-0)\,P$_{1}$+18.5 & 
16310.142  & 1.6814 &   -5.7275 &   214    & (6-3)\,R62 \\ 
17109.3710 & 0.96  &   -4.64  &     409   & (3-1)\,P$_{1}$-15.5 & 
16310.965  & 2.0492 &   -5.8204 &   165    & (5-2)\,R81 \\ 
17111.8320 & 0.84  &   -5.01  &     427   & (2-0)\,P$_{1}$+19.5 & 
16312.011  & 0.7920&  -7.1364 &   203    & (6-3)\,R 4 \\ 
17114.9355 & 0.96  &   -4.64  &     446   & (3-1)\,P$_{1}$+15.5 & 
23091.567  & 1.4401&   -4.9234&   373    & (2-0)\,R78 \\ 
17320.8769 & 1.03  &   -4.59  &     442   & (3-1)\,P$_{1}$+16.5 & 
23103.397  & 1.4763&   -4.9147&   355    & (2-0)\,R79 \\ 
17338.7148 & 0.93  &   -4.97  &     379   & (2-0)\,P$_{1}$+19.5 & 
23115.676  & 1.5128&   -4.9062&   305    & (2-0)\,R80 \\ 
17590.4589 & 1.01  &   -4.92  &     405   & (2-0)\,P$_{1}$-20.5 & 
23128.407  & 1.5499&   -4.8979&   318    & (2-0)\,R81 \\ 
17594.9785 & 1.01  &   -4.92  &     456   & (2-0)\,P$_{1}$-21.5 & 
HF         &       &          &           &             \\ 
17597.3300 & 1.01  &   -4.92  &     310   & (2-0)\,P$_{1}$+20.5 & 
23364.7589 &  0.48 &   -3.9547&   514    & (1-0)\,R 9 \\ 
17603.7734 & 1.01  &   -4.92  &     335   & (2-0)\,P$_{1}$+21.5 & 
22964.2745 &  0.71 &   -3.9393&   332    & (1-0)\,R13 \\ 
17612.1152 & 1.24  &   -4.42  &     289   & (4-2)\,P$_{2}$-12. & 
22893.0391 &  0.78 &   -3.9431&   334    & (1-0)\,R14 \\ 
17623.7226 & 1.24  &   -4.42  &     300   & (4-2)\,P$_{1}$-13.5 & 
22784.5448 &  0.93 &   -3.9626&   265    & (1-0)\,R16 \\ 
CN         &       &          &           &           & 
$^{13}$C$^{16}$O & &          &          &             \\ 
21911.0848 & 1.13  &   -1.87  &     86   & (0-2)\,Q$_{1}$57 & 
23655.522    &  0.09   &   -5.73   &   584    & (2-0)\,R19 \\ 
21934.7310 & 0.89  &   -2.28  &     54   & (0-2)\,P$_{1}$47 & 
23605.117    &  1.45   &   -4.94   &   245    & (2-0)\,R80 \\ 
21941.1191 & 0.89  &   -2.14  &    79   & (1-3)\,P$_{1}$34 & 
23595.911    &  0.14   &   -5.62   &   514    & (2-0)\,R24 \\ 
21942.6244 & 0.91  &   -2.13  &     85   & (1-3)\,P$_{2}$35 & 
23486.772    &  0.32   &   -5.40   &   564    & (2-0)\,R37 \\ 
21945.2729 & 1.07  &   -1.69  &    104   & (0-2)\,Q$_{1}$44 & 
23475.782    &  0.35   &   -5.37   &   457    & (2-0)\,R39 \\ 
21895.4928 & 1.07  &   -1.71  &    108   & (0-2)\,Q$_{2}$44 & 
23473.000    &  0.94   &   -5.08   &   387    & (2-0)\,R64 \\ 
21923.8628 & 1.37  &   -1.88  &     53   & (2-4)\,R$_{2}$57 & 
23470.918    &  0.37   &   -5.36   &   492    & (2-0)\,R40 \\ 
21956.4998 & 0.94  &   -2.37  &     59   & (2-4)\,P$_{2}$18 & 
23466.497    &  0.39   &   -5.34   &   462    & (2-0)\,R41 \\ 
21914.6381 & 0.99  &   -1.89  &     82   & (2-4)\,Q$_{1}$23 & 
23460.072    &  0.85   &   -5.11   &   380    & (2-0)\,R61 \\ 
21939.3027 & 1.02  &   -1.85  &     85   & (2-4)\,Q$_{2}$26 & 
$^{12}$C$^{17}$O&  &          &          &                 \\ 
21943.5874 & 1.20  &   -1.94  &     94   & (2-4)\,R$_{1}$38 & 
23358.509    &  0.15   &   -5.58   &   162    & (2-0)\,R25 \\ 
21926.9394 & 1.22  &   -1.94  &     75   & (2-4)\,R$_{2}$39 & 
23338.271    &  0.18   &   -5.54   &   174    & (2-0)\,R27 \\ 
CO         &       &          &          &                  & 
23311.069    &  0.22   &   -5.49   &   207    & (2-0)\,R30 \\ 
16163.720  & 0.3464 & -7.4573 &   257    & (4-1)\,P18 & 
23328.945    &  0.19   &   -5.53   &   164    & (2-0)\,R28 \\ 
16178.560  & 0.3554 & -7.4383 &   220    & (4-1)\,P19 & 
23319.865    &  0.20   &   -5.51   &   193    & (2-0)\,R29 \\ 
16266.752  & 0.8128 & -6.7650 &   286    & (6-3)\,R10 & 
23302.812    &  0.23   &   -5.48   &   142    & (2-0)\,R31 \\ 
16278.366  & 1.5718 & -5.7791 &   261    & (6-3)\,R58 & 
23295.035    &  0.24   &   -5.46   &   147    & (2-0)\,R32 \\ 
16280.684  & 0.8040 & -6.8620 &   226    & (6-3)\,R 8 & 
23287.433    &  0.26   &   -5.44   &   144    & (2-0)\,R33 \\ 
\hline 
\end{tabular} 
 
\end{table*} 

\section{Analysis and results} 
The standard LTE line analysis program WIDTH9 developed by R.~L.~Kurucz
has been employed for abundance calculations of atomic lines.   Model
atmospheres were extracted from Hauschildt et al.
(\cite{hauschildt99}).  The most recent version of the spectrum
synthesis code MOOG (Sneden \cite{sneden}) was used for analysis of
selected molecular lines. Spectral lines stronger than 300 m\AA\ were
not used in general for abundance calculations.  The abundances using
the same sample of lines were calculated for CH~Cyg and the Sun to
detect possible systematic effects. For solar atmosphere the Kurucz's
model (Kurucz \cite{kurucz93}) of the Sun was adopted.  The
calculations of synthetic spectra over the entire observed spectral
region were done using the code WIDMO developed by one of us (M.S.).

\subsection{Atmospheric parameters} 
The effective temperature of CH~Cyg was estimated by Dyck, Belle, \&
Thompson (\cite{dyck99}) using interferometric measurements of the
angular diameter, $T_\mathrm{eff} = 3084 \pm 130\, \rm K$. Effective
temperature as a function of spectral type for M giants was calibrated
by Richichi et al. (\cite{richichi99}) using lunar occultation
measurements corrected by inclusion limb darkening effects. According
to this calibration spectral type M7 corresponds to an effective
temperature of 3150$\pm$95~K. Photometric estimation of temperature is
rather difficult for CH~Cyg because of the variability of colour
indexes.  The temperature $T_\mathrm{eff}$ = 3100~K seems to be good
approximation for this cool giant.

The surface gravity ($\log g$) for CH~Cyg is not possible to 
determine in the standard way from the \ion{Fe}{i}/\ion{Fe}{ii} 
ionization balance due to absence of detectable lines of ions 
in the cool atmosphere of giant. Molecular lines are not useful 
for this task because partial pressures are not sensitive to the  
total pressure. Surface gravity may be found 
from the trigonometric parallax, $\piup = 3.73 \pm 0.85$ mas (ESA 1997). 
Using this value, $m_\mathrm{V}$  from 7.40 to 9.10 mag (SIMBAD), 
$BC = -4.42$, $T_\mathrm{eff} = 3100$~K and adopting a mass of 
$M = 2\, \rm M_{\sun}$ the standard relation\\  
$\log g = 4 \log T_\mathrm{eff}/T_{\sun} + \log M/M_{\sun} + 2 \log \piup + 0.4(V+BC-0.26) + 4.44$ \\ 
leads to $\log g$ from -0.1 to +0.6 (cgs). This estimation is close to 
the standard value of 0.0 for M7 giants (Houdashelt et al. \cite{Houdashelt}).  

The microturbulent velocity for CH~Cyg was estimated by forcing the
abundances of the individual \ion{Fe}{i} lines to be independent of the
equivalent widths, $\xi_t$ = 2.2~km\,s$^{-1}$.  The macroturbulent
velocity was introduced in order to fit the profiles of synthesized and
observed lines. The FWHM value of 5.9~km\,s$^{-1}$ was measured using a
number of weak symmetric lines. Notice that Hinkle et al.
(\cite{hinkle93}) found a variability of FWHM during the short-period
orbital motion.

The atmospheric parameters adopted for abundance calculations of CH~Cyg
are as follows: $T_{\rm eff} = 3200$~K, $\log g$ = 0.0 (cgs), and
$\xi_t$ = 2.2~km\,s$^{-1}$, nearest in the grid of model atmospheres
(Hauschildt et al. \cite{hauschildt99}) and close to the mean values
for single M giants (Houdashelt et al. \cite{Houdashelt}).
Uncertainties in the adopted atmospheric parameters were estimated to
be $\pm$ 200~K in temperature, $\pm$0.5~dex in gravity and
$\pm$0.5~km\,s$^{-1}$ in microturbulent velocity. For the Sun standard
atmospheric parameters (5777~K, 4.44 (cgs), 1.0~km\,s$^{-1}$) were
accepted.

\subsection{Atomic and Molecular data} 

The insufficient quality of available atomic and molecular data is a
serious limitation for abundance analysis using lines in the infrared
spectral region. The abundances of Mg, Si, Ti, Fe, and Ni in this paper
were calculated using atomic lines. The atomic data for measured lines
were collected using the VALD database (Kupka et al. \cite{kupka99})
and the list prepared by M\'elendez \& Barbuy (\cite{melendez99}).  The
quality of the collected oscillator strengths was inspected via
abundance calculations of the Sun. Unfortunately, not all lines
selected and measured in the spectrum of CH~Cyg are detectable in the
spectrum of Sun.  The solar abundance of \ion{Fe}{i} was calculated
using lines in the CH~Cyg line list.  Abundances from individual lines
were compared to the mean and lines giving incorrect values were
omitted.  An average metallicity of the Sun was found to be
7.54$\pm$0.12 (see Table 3) from the best lines, close to that obtained
in the optical region, 7.50$\pm$0.12 (Grevesse \& Sauval
\cite{grevesse}). Then the same sample of lines with the same ($\log
gf$) was used to calculate the iron abundance of CH~Cyg.  This
minimized the role of systematic effects (see the last column in Table
3).

The same methodology was used to select lines of \ion{Mg}{i},
\ion{Si}{i}, \ion{Ti}{i}, and \ion{Ni}{i} with best available
oscillator strengths. The solar abundances of these elements,
calculated using these lines, are presented in Table 3.  The systematic
effects and standard deviations in abundances mainly due to errors in
$\log gf$ are in general lower than those due to uncertainties in the
atmospheric parameters for CH~Cyg (see Table 4). Unfortunately, our
efforts to create a sample of lines with acceptable oscillator
strengths for \ion{Na}{i}, \ion{Al}{i}, \ion{S}{i}, \ion{Ca}{i},
\ion{Sc}{i}, \ion{V}{i}, \ion{Cr}{i} was unsuccessful mostly due to
serious problems with quality of oscillator strengths for these
elements in the infrared.

The abundances of CNO elements and fluorine were calculated using the
molecular lines.  The vacuum wavelengths, excitation potentials, and
$gf$-values for the vibration-rotation lines of CO isotopes are adopted
from Goorvitch (\cite{goorvitch2}).  Both $\Delta v$\,=\,2 second
overtone lines in K-band and $\Delta v$\,=\,3 third overtone lines in
H-band were involved in the analysis.  The accepted dissociation energy
of CO was D$_{0}$=11.091~eV.  In the case of $^{12}$CO, $^{13}$CO,
$^{12}$CN, and OH lines the list of lines used by Smith \& Lambert
(1990) was adopted.  The abundance of fluorine was calculated using
rotation-vibration lines of HF molecule.  HF data given by Jorissen et
al. (\cite{jorissen92}) were used.

\subsection{Abundances} 

The mean absolute and relative abundances in the scale of $\log
\varepsilon$(H) = 12.0 derived with $T_{\rm eff}$ = 3200 K, $\log$g =
0.0, $\xi_{\rm t}$ = 2.2~km\,s$^{-1}$ and [M/H] = 0.0 for CH~Cyg are
given in Table 3, together with the standard deviations of abundances
estimated from individual lines, and the number (n) of lines used in
the analysis. The abundances relative to the Sun ([X]) were calculated
using solar photospheric data provided by Grevesse \& Sauval
\cite{grevesse}, except in the case of fluorine, for which we used the
meteoritic solar abundance of 4.48.  The systematic errors in
abundances produced by uncertainties in $T_{\rm eff}$ ($\pm$200~K),
$\log g$ ($\pm$0.5~dex), and $\xi_t$ ($\pm$0.5~km\,s$^{-1}$) would lead
to errors, less than 0.3 dex for all elements (see Table 4).

\begin{table} 
 
\caption[]{Averaged absolute and relative abundances for CH~Cyg and the Sun calculated using the same sample of absorption lines. 
The standard deviations and the number of lines used in the analysis 
are also given.} 
\label{TabAbund} 
\begin{tabular}{ccccccc} 
\hline 
   & CH~Cyg &  &     & Sun  &    &     \\ 
X & $\log \epsilon$(X)& n & [X] & $\log \epsilon$(X)& n & [X] \\ 
\hline 
C  & 8.37$\pm$0.22 & 11 & $-$0.15&             &    &       \\ 
N  & 8.08$\pm$0.13 & 13 & +0.16  &               &    &       \\ 
O  & 8.76$\pm$0.24 & 15 & $-$0.07&             &    &       \\ 
F  & 4.65$\pm$0.15 &  4 & +0.17  &               &    &       \\ 
Mg & 8.68$\pm$0.21 & 10 & +1.10  & 7.40$ \pm$0.12 &  4 & $-$0.18 \\ 
Si & 7.40$\pm$0.18 & 13 & $-$0.15  & 7.43$ \pm$0.24 & 11 & $-$0.12 \\ 
Ti & 5.24$\pm$0.09 &  7 & +0.22  & 5.05$ \pm$0.19 &  6 & +0.03 \\ 
Fe & 7.50$\pm$0.19 & 51 & +0.00  & 7.54$\pm$0.12  & 51 & +0.04 \\ 
Ni & 6.65$\pm$0.19 & 12 & +0.40  & 6.36$\pm$0.22  & 10 & +0.11 \\ 
\hline 
\end{tabular} 
\end{table} 

\begin{figure*} 
\resizebox{\hsize}{!}{\includegraphics{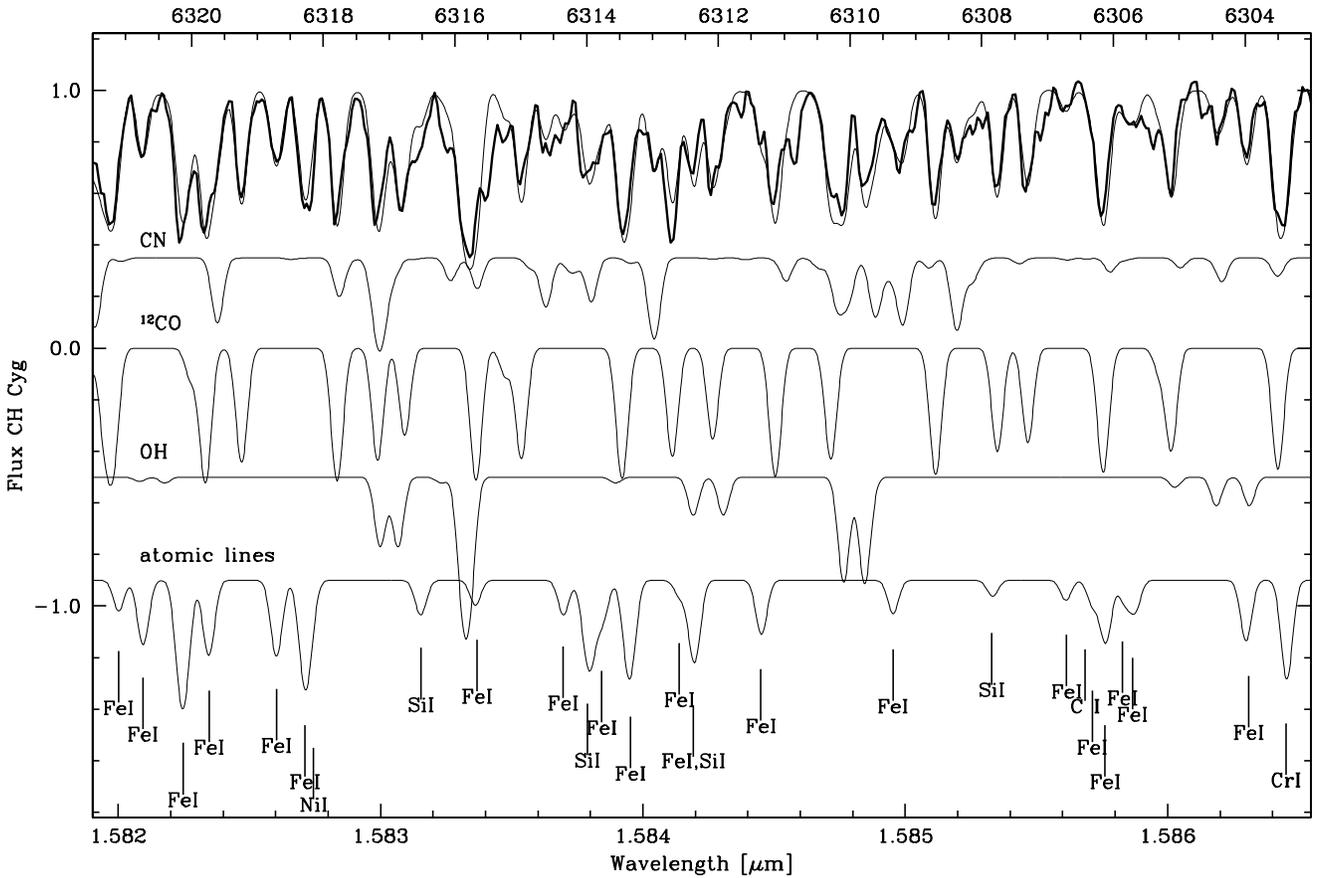}} 
\caption{The observed (thick line) and synthetic spectra calculated using final abundances of the 
symbiotic star CH~Cyg in the spectral region from 1.582 to 1.586 $\muup$m are shown on the top. The synthesized CN, OH, $^{12}$CO, and atomic components are presented separately. Atomic lines are identified by ticks.} 
\label{FigFe} 
\end{figure*} 

\begin{table} 
\caption{Sensitivity of abundances to uncertainties in stellar parameters.} 
\label{TabSens} 
\begin{tabular}{cccc} 
\hline 
$\Delta\,X$ & $\Delta T_\mathrm{eff} = -200\,\mathrm{K}$ & $\Delta 
\log g=+0.50$ & $\Delta \xi_t = +0.5$ \\ 
\hline 
C  & $-$0.11 &  +0.20  & $-$0.11 \\ 
N  & $-$0.13 & $-$0.25 & $-$0.06 \\ 
O  & $-$0.26 & $-$0.01 & $-$0.15 \\ 
F  & $-$0.30 &  +0.18  & $-$0.13 \\ 
Ti & $-$0.20 &  +0.10  & $-$0.30 \\ 
Fe & $ $0.07 &  +0.13  & $-$0.12 \\ 
\hline 
\end{tabular} 
\end{table} 

\subsection{Isotopic ratios} 

The high resolution spectra provide an opportunity to 
estimate the isotopic ratios of carbon and oxygen using individual  
lines of $^{13}$C$^{16}$O, $^{12}$C$^{17}$O and $^{12}$C$^{18}$O. 
The spectral region redward of the $^{13}$CO (2-0) band head 
is significantly contaminated by telluric lines. 
However, nine clean $^{13}$C$^{16}$O lines (Table 2) were selected  
between the band heads of the systems (2-0) $^{13}$C$^{16}$O and (4-2)  
of $^{12}$C$^{16}$O. The ratio $^{12}$C/$^{13}$C was calculated to be 
18$^{+12}_{-6}$. 

A relatively uncontaminated sample of $^{12}$C$^{17}$O  
lines was identified in the spectral range from 4278 to 4294 cm$^{-1}$.  
This region also has a number of strong telluric lines. To check 
clean profiles the spectrum was rationed by the 
logarithmically corrected telluric spectrum from the Arcturus Atlas.    
In addition, the synthetic spectrum was calculated over the entire range  
to see possible contamination by other molecular ($^{12}$C$^{16}$O, CN) lines
in the stellar spectrum. 
Eight clean $^{12}$C$^{17}$O lines were selected and  
measured (see Table 2) and the $^{16}$O/$^{17}$O ratio was estimated  
to be 830$^{+400}_{-270}$.     

Due to the complex telluric spectrum the selection of clean lines 
in the region of (2-0) $^{12}$C$^{18}$O band head was problematic. A  
crude estimate of the $^{16}$O/$^{18}$O isotopic ratio was made using 
the spectral synthesis method, $^{16}$O/$^{18}$O $>$ 1000.   

\section{Discussion} 

An LTE abundance analysis of the red symbiotic star CH~Cyg was made for
the first time using high-resolution near-infrared spectra and
atmospheric models.  Weak and medium strong (EW$<$300 m\AA) atomic
lines formed deeply in the atmosphere of cool giant were used
to calculate abundances of iron and $\alpha$-group elements. The
abundance of iron was found to be solar.  The mean of three selected iron
group elements (Ti, Fe, Ni) displays slight overabundance, [M] = +0.2 dex. 
Possible uncertainties in atmospheric
parameters lead to errors of no more than 0.3 dex for all elements.

CNO abundances were determined using first and second
overtone CO lines, first overtone OH lines, and CN red system lines.  [C/H] = -0.15, [N/H] = +0.16 and
[O/H] = -0.07, close to the averaged values obtained for single M
giants by Smith \& Lambert (\cite{smith90}).  C/O = 0.40 again in agreement
with bright field M giants.   The
isotopic ratios $^{12}$C/$^{13}$C = 18  and $^{16}$O/$^{17}$O = 830
calculated using isotopic absorption lines also are approximately equal
to those for single M giants.

In the standard picture of a symbiotic binary the hot WD
component ionizes the cool giant wind giving rise to strong UV emission
lines of \ion{C}{iii}], \ion{C}{iv}, \ion{N}{iii}], {\ion{N}{iv}] and
\ion{O}{iii}]. Based on emission line
fluxes from these ions, Nussbaumer et al. (\cite{nussbaumer88}) deduced C/N and O/N
abundance ratios for 24 symbiotic systems observed with the {\it IUE}
satellite. They found that the CNO abundance ratios place the symbiotic
stars in the transition region from giants to supergiants, almost
coinciding with the M-giants. They also noted that their method tends
to overestimate nitrogen relative to oxygen and carbon, although the
effect is not larger than $\sim 30\, \%$.  

Our abundances calculated
from the M giant absorption lines can be compared with those derived
from the UV emission lines providing an opportunity for a
verification of the method used by Nussbaumer et al.
(\cite{nussbaumer88}), and more generally any method based on emission
lines.  The fluxes of emission lines for CH~Cyg were measured using low
resolution {\it IUE} observations obtained during 1979-1986 by 
Miko\l ajewska et al. (\cite{mikolajewska88}), to which we added
measurements from {\it IUE} observations through May 1989. The relative
abundance of C/N and O/N were determined using 22 observations
following the procedure described by Nussbaumer et al.
(\cite{nussbaumer88}). The average ratios were: C/N = 0.57, O/N = 2.0,
and C/O = 0.28.  The ratios place CH~Cyg among other symbiotic systems in
the C/N--O/N plane (cf. Figs. 2--4 of Nussbaumer et al. 1988).  However, the
standard deviations of the mean values of C/N and O/N are about
$\sim$50$\%$. The standard deviation of the mean C/O ratio is lower,
$\sim$28$\%$. It seems that the fluxes of oxygen and carbon emission
lines are changing in phase, while nitrogen emission lines evolved
differently. 

The mean C/O ratio obtained using emission line techniques
is close to that calculated using molecular absorption lines, C/O =
0.4. However, the differences in C/N and O/N ratios between two methods
are significant. The molecular absorption lines give higher values,
C/N=1.6 and O/N=4.0, respectively, and on the C/N--O/N plane place 
CH~Cyg among M giants. 
An important source of uncertainties in the photospheric abundance analysis is the dissociation energy of CN, D$^0_0$(CN).
For this study, we have adopted D$^0_0$(CN)=7.65~eV from Bauschlicher et al. (\cite{bauschlicher88}). 
Adopting a higher value of D$^0_0$(CN)=7.77~eV (Costes et al. \cite{costes90}) would reduce nitrogen 
abundance by 0.28 dex increasing even more the discrepancy with the emission line results.
The N abundance could be increased by adopting a
lower value of the surface gravity (see Table 4). However,
with the available grid of models,  a quantitative analysis of this hypothesis is not yet possible . 
As long as we assumed correctness of the determination of the nitrogen abudances, this result suggests that at least in case of 
CH~Cyg, the method based on emission lines may overestimate the abundance
of nitrogen by a factor of a few. Alternatively, it is possible that in
CH~Cyg, the emission lines originate in the ejecta/jets from the hot
component rather than in the illuminated parts of the M giant wind.

The atlas of spectra of cool stars published by Wallace \& Hinkle
(\cite{wallace96}) provides spectra of two M giants - $\lambda$~Dra
(M0\,III) and RX~Boo (M8\,III) and two M supergiants - $\alpha$~Ori
(M2\,Ia) and $\alpha$~Her (M5\,Ib-II). Qualitative comparison of
these spectra shows differences among the spectra of single M
(super)giants and the cool giant of CH~Cyg. In the spectra of single
(super)giants the depth of (2-0) $^{12}$C$^{16}$O band head is about 65
- 70$\%$, whereas in CH~Cyg the CO band heads are never deeper than
60$\%$ (Fig.4). The same conclusion may be drawn inspecting the strong
vibrational-rotational lines of CO bands in the K band. A quantitative
comparison of the equivalent widths of (2-0) high rotational
transitions lines R78-R81 with the data for some M giants (Smith \&
Lambert \cite{smith90}) shows that the equivalent widths
of these lines measured in CH~Cyg (M7) are typical for early M stars.
A speculation is that the differences in the CO lines results from the upper
atmosphere of CH~Cyg cool giant being modified by $X$-$UV$-radiation of WD
companion. 
However, the spectra chosen for the analysis were taken when the red giant was closer to us than the hot component according to the long period orbit of Hinkle et al.\cite{hinkle93}, and the illuminated hemisphere could not be well visible. Moreover, the depth of (2-0) $^{12}$C$^{16}$O band head reached the lowest value in 1987-88 when the hot component luminosity reached its minimum value (e.g. Miko{\l}ajewska, Miko{\l}ajewski \& Selvelli \cite{jmik93}) which could be hardly reconciled with the illumination hypothesis.
A caveat is that weakening of the strongest lines has also 
been seen in spectra of late-type large amplitude variable stars, perhaps 
as the result of either extended atmospheric layers or
mechanical energy transport to the more tenuous regions of the atmosphere, or
a combination of these effects (Hinkle, Hall \& Ridgway \cite{hhr82}).  

Another possibility is the presence of the hot dust envelope.  In
fact, ISO spectra do show the existence of an extended dust envelope
around CH~Cyg (Schild et al. \cite{schild99}). Calculations with the DUSTY code
(Ivezi{\'c} \& Elitzur \cite{ivezic95}, \cite{ivezic97}) show that the contribution of the
flux from the dust to the total flux at K band wavelengths can reach at
most a few percent.  So this possibility cannot be totally rejected.
Perhaps this is supported by the fact that CO bands in H band are well
reproduced by the synthetic spectrum, and that minimum band depth is coincinding with the maximum of the $M$ band light curve of Munari et al. (\cite{munari96}).  However, elaboration on this
topic is outside the scope of the paper.

\begin{figure} 
  \resizebox{\hsize}{!}{\includegraphics{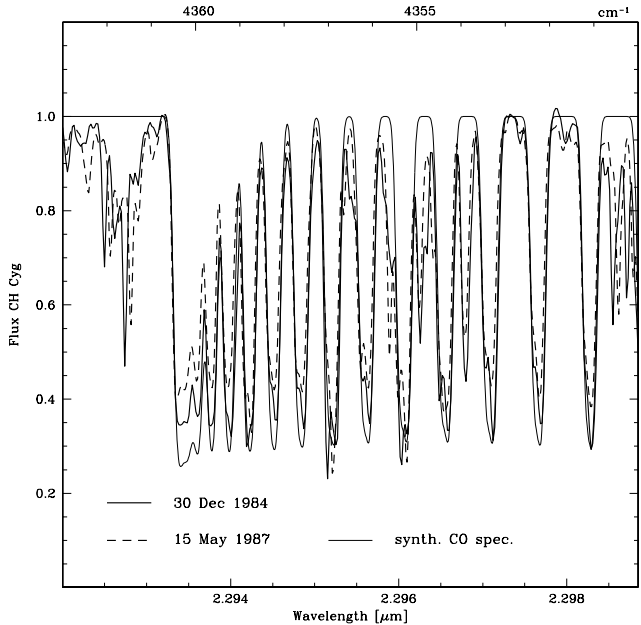}} 
  \caption{The observed spectra  (thick solid an dashed) of CH~Cyg in the region
of (2-0) $^{12}$C$^{16}$O band head for two different phases of
symbiotic activity. Also shown is the calculated spectrum (thin solid) using
the final atmospheric parameters and abundances.} 
  \label{fig4} 
\end{figure} 

To fit the equivalent widths of the lines
(2-0) R78-R81 the carbon abundance of CH~Cyg would have to be changed by
$-$0.7 dex, in conflict with the carbon abundance calculated
from the weak second overtone CO lines. The reason can not to be an
error in the effective temperature adopted because we would have to raise the
temperature to 3800~K to fit R78-R81 equivalent widths. Such a high
effective temperature is inconsistent with the spectroscopic type and 
colours.
Smith \& Lambert (\cite{smith90}) discussed in details similar effects
for single M giants and concluded that strongest CO lines are not
simple monitors of carbon abundance. These lines should be rejected in
deriving both the microturbulence and the abundance because the
outermost layers where strong lines are formed differ from the
model atmosphere.  Mechanical heating may lead to higher
temperatures and/or microturbulence in the surface layers.

\section{Conclusions} 

1. The photospheric abundances for the cool component of symbiotic star
CH~Cyg were calculated for the first time using high-resolution,
near-infrared spectra and the method of standard LTE analysis and
atmospheric models.  Iron abundance for CH~Cyg was found to be solar,
[Fe/H] = +0.0$\pm$0.19.  The atmospheric parameters, ($T_{\rm eff}$ =
3200~K, $\log g = 0.0$ (cgs), $\xi_{\rm t} = 2.2$~km\,s$^{-1}$) and
metallicity, for CH~Cyg are approximately equal to those for normal M7
giants.

2. Although the orbital period of symbiotic system CH~Cyg of $\sim 5300^\mathrm{d}$ is not unusual
as compared with orbital periods of barium stars (Jorissen et al. \cite{jorissen98})\footnote{Note that different chanels of mass transfer in binaries can produce barium (peculiar) stars with orbital periods from 10 to 400 000 days
(Han et al. \cite{han95}).},
CH~Cyg does not display any abundance peculiarities in
comparison with single field M giants. 
The calculated [C/H] = -0.15,
[N/H] = +0.16, and [O/H] = -0.07 are close to the averaged values for
six M giants, [C/Fe] = -0.14$\pm$0.09, [N/Fe] = +0.44$\pm$0.13 and
[O/Fe] = -0.08$\pm$0.08 analyzed by Smith \& Lambert (\cite{smith90}).
The isotopic ratios of $^{12}$C/$^{13}$C and $^{16}$O/$^{17}$O for
CH~Cyg are found to be 18$^{+12}_{-6}$ and 830$^{+400}_{-270}$,
respectively, in agreement with those for giants experiencing the first
dredge-up. There is no signature of significant $s$-process
enhancement, so synthesis of neutron-capture elements in
WD progenitor (AGB star in the past) or mass transfer to the secondary
(now primary) was not efficient. The reason of this could be high
metallicity of the star, in agreement with the scenario proposed by
Jorissen (\cite{jorissen03}; and references therein). By comparison the
$yellow$ symbiotic stars, which have significant enhancement 
of s-process elements, are metal-deficient.

3. The C/O ratio calculated for CH~Cyg using the technique of emission
lines provided by Nussbaumer et al. (\cite{nussbaumer88}) is close to
that obtained from analysis of the absorption lines.  However, results
for C/N and O/N ratios derived from the two techniques are significantly
different. 

4. The strong lines in the CH~Cyg infrared spectrum are more shallow 
than those of standard M giants.   This may reflect modification by
high-energy radiation of the WD companion.  In any case, abundances
derived using strong ($>$300 m\AA) lines are sensitive to the
atmospheric model. 

\begin{acknowledgements} 
This research was partly founded by KBN Research Grants Nos. 
5\,P03D\,019\,20, and 1\,P03D\,017\,27. 
The collaborative program between Polish and Latvian Academy of 
Sciences is acknowledged for support. We thank the referee, Hans Martin Schmid, for important sugestions which improved the paper.
\end{acknowledgements}

\end{document}